\documentclass[journal, final]{IEEEtran}

\usepackage[pdftex]{graphicx}
\pdfoutput=1
\ifCLASSINFOpdf
\else
\fi
\usepackage{xcolor} 
\usepackage{amsmath}
\usepackage{algorithm} 
\usepackage{algpseudocode} 
\usepackage[hyphens]{url}
\usepackage{subfigure}
\usepackage[american,cuteinductors,smartlabels]{circuitikz}
\usepackage{siunitx}
\usepackage{amsmath,amssymb}
\usepackage{amsthm} 
\usepackage[mode=buildnew]{standalone}
\usepackage{mathtools}
\usepackage{multirow}
\allowdisplaybreaks[4]

\begin{document}
%
\title{Acceleration of Power System Dynamic Simulations using a Deep Equilibrium \\ Layer and Neural ODE Surrogate}
%
%

\author{Matthew~Bossart,~\IEEEmembership{Student Member,~IEEE,}
        Jose Daniel Lara,~\IEEEmembership{Senior Member,~IEEE,}
        Ciaran~Roberts,~\IEEEmembership{Member,~IEEE,}
        Rodrigo Henriquez-Auba,~\IEEEmembership{Member,~IEEE,} 
        Duncan Callaway,~\IEEEmembership{Member,~IEEE,}
        and~Bri-Mathias~Hodge,~\IEEEmembership{Senior Member,~IEEE}
}

\maketitle

\begin{abstract}
The dominant paradigm for power system dynamic simulation is to build system-level simulations by combining physics-based models of individual components. The sheer size of the system along with the rapid integration of inverter-based resources exacerbates the computational burden of running time domain simulations. In this paper, we propose a data-driven surrogate model based on implicit machine learning---specifically deep equilibrium layers and neural ordinary differential equations---to learn a reduced order model of a portion of the full underlying system. The data-driven surrogate achieves similar accuracy and reduction in simulation time compared to a physics-based surrogate, without the constraint of requiring detailed knowledge of the underlying dynamic models. This work also establishes key requirements needed to integrate the surrogate into existing simulation workflows; the proposed surrogate is initialized to a steady state operating point that matches the power flow solution by design. 

\end{abstract}


%
\IEEEpeerreviewmaketitle

\section{Introduction}
%
%
%
%

\IEEEPARstart{T}{he} transition from fossil fuel-based generation to renewable energy sources is accelerating. Unlike traditional generation based on rotating synchronous machines, wind, solar photovoltaics, and storage are inverter-based resources (IBRs) that interface with the grid through power electronics. Synchronous machines and IBRs have fundamentally different dynamic behavior, and therefore the adoption of IBRs has profound impacts on the dynamic behavior of power systems.   
 
Given the fundamental shift in the dynamics of the system brought on by IBRs, it is not guaranteed that historically useful frameworks for the study of system stability will provide adequate results for future systems. Recent work has shown that including line dynamics can change system-wide stability classification \cite{markovic_stability_2018, lara_revisiting_2023, henriquez2020grid, brouillon_effect_2018} and that it is difficult to accurately capture fast inverter controls, such as the inner current loop, in the positive sequence domain \cite{wang_instability_2018}. Events in Australia and Texas demonstrate the limitations of positive sequence tools to capture observed system oscillations \cite{badrzadeh_electromagnetic_nodate, noauthor_odessa_2021}. 

These recent challenges in IBR simulations highlight the need for more detailed modeling (i.e. more dynamic states) which in many cases includes the representation of Electromagnetic Transients (EMT). Furthermore, the combination of  EMT dynamics and the fast control loops of IBRs imposes a smaller time step requirement for solving systems with IBRs \cite{kenyon_comparison_2023}. Considering that the time to solve systems of differential equations is largely determined by (1)~the number of states in the system and (2)~the time step required for solving the system \cite{hairer_solving_1996}, simulating systems with large penetrations of IBRs will be more computationally expensive. For a more in-depth discussion on the underlying assumptions of various time domain simulation approaches, refer to \cite{lara_revisiting_2023_tpwrs}. The need for increased detail in simulation models is at odds with the need to carry out studies that require simulating multiple scenarios, such as contingency analysis or system impact studies \cite{charles2023off}. 

\begin{figure}[t]
    \centering
    \includegraphics[width=0.48\textwidth]{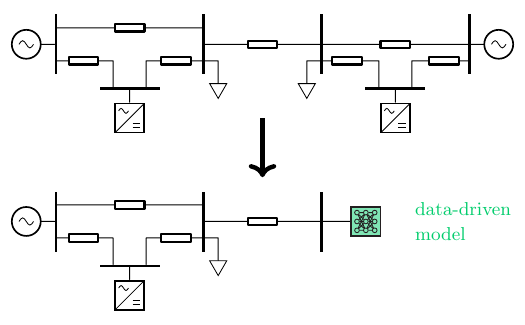}
    \caption{For large systems, modeling all devices in detail (above) becomes computationally intractable. This paper proposes a data-driven surrogate model that is designed to be integrated into simulations containing physics-based models (below).}
    \label{fig:intro}
\end{figure}

A useful technique to accelerate simulations is to create simpler versions of complex models called ``reduced-order models" or ``surrogates". These surrogates can capture intricate system dynamics with fewer states, significantly reducing the computational demands during simulation. In this manuscript, we use the term ``physics-based surrogate model" to describe a surrogate with a specified structure that is designed to replicate specific observed dynamics. Prior works have developed ``physics-based surrogate models" by representing many devices with a single aggregate model of similar structure \cite{purba_reduced-order_2019,purba_dynamic_2020, ma_mathematical_2020}. Developing these surrogate models requires knowledge of the constituent models' structure and parameters, which is not always available. For instance, simulations that involve manufacturer-provided black box models will not have the internal structure revealed to the model user. In addition, power systems with inverter control and distributed energy resources (DERs) can demonstrate complex dynamic behaviors that a structured model may not fully encapsulate in a general sense. In this case, the physics-based surrogates must rely on distinct parameterizations in order to capture the dynamic behavior accurately for different simulations \cite{noauthor_model_2018, noauthor_value_2015}. These drawbacks can be mitigated by employing other types of surrogate models that are capable of emulating the underlying system's behavior without relying on structural assumptions.

The ``data-driven surrogate model" presented in this paper is capable of creating a reduced-order representation with no underlying structural assumptions (Fig. \ref{fig:intro}). Unlike physics-based surrogate models that require extensive design and validation, data-driven surrogates can be trained on simulation data to capture the dynamics of a sub-section of a system. These surrogates are adequate for modeling complex, non-linear behavior, and the number of states can be adjusted to balance computational cost and accuracy, based on the complexity of the underlying system. The framework also captures faster time-scale phenomena, such as network dynamics and inner control loops of IBRs, when necessary. Data-driven surrogates have proven to be effective in accelerating simulations in other fields, such as chemical reactions and heating systems \cite{anantharaman_accelerating_2021}.

\begin{figure}[t]
     \centering
     \subfigure[Direct classification of stability \protect\cite{ren_interpretable_2022, sarajcev_power_2021, senyuk_power_2023}.]
     {
         \centering
         \includegraphics[width=0.45\textwidth]{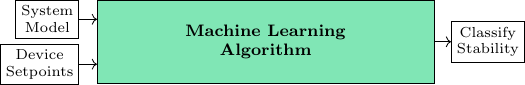}
         \label{fig:ML_A} 
    }
     \subfigure[Predicting time domain trajectories (e.g. physics-informed neural networks) \protect\cite{misyris_physics-informed_2020}.]
         {
         \centering
         \includegraphics[width=0.45\textwidth]{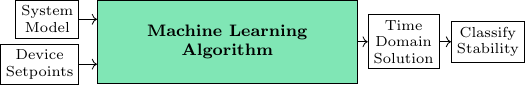}

         \label{fig:ML_B} 
         }
     \subfigure[Incorporating ML models within the existing transient simulation workflow \protect\cite{vorwerk_using_2022, azmy_identification_2004, zhou_neuro-reachability_2022, xiao_feasibility_2022}.]
         {
         \centering
         \includegraphics[width=0.45\textwidth]{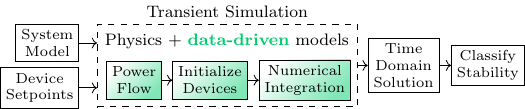}
         \label{fig:ML_C}
         }
        \caption{Types of applications of ML to power system dynamics.}
        \label{fig:three graphs}
\end{figure}

In previous applications of scientific machine learning (ML) to power system dynamic simulation, the focus was on completely replacing the transient simulation process. This involved either classifying stability (Fig. \ref{fig:ML_A}) or predicting the time domain solution directly (Fig. \ref{fig:ML_B}). We take a different approach by integrating data-driven models into an existing transient simulation workflow (Fig. \ref{fig:ML_C}). Prior works that follow a similar approach have developed data-driven models based on variations of recurrent neural networks to predict the current state based on a fixed sequence of prior states \cite{vorwerk_using_2022, azmy_identification_2004}. These surrogates are trained based on prior values recorded at a fixed simulation time step; integrating them into a larger system therefore eliminates the possibility of using adaptive time-stepping. A more natural choice for seamless integration is a neural ordinary differential equation (neural ODE), which can be seen as an extension of recurrent neural networks to continuous time \cite{chen_neural_nodate}. The authors in \cite{zhou_neuro-reachability_2022} use neural ODEs to model the uncertainty of DERs in systems of networked microgrids; in \cite{xiao_feasibility_2022}, neural ODEs are used for parameter estimation and black box modeling. However, neither of these works addresses essential considerations related to integrating data-driven models (i.e., End-to-End implementation), such as initializing the models to start in steady-state and respecting the boundary conditions imposed by physics-based devices and the power flow solution. 

The novel surrogate structure proposed in this paper is based on deep equilibrium (DEQ) layers \cite{bai_deep_nodate} and neural ODEs \cite{chen_neural_nodate}, which are both examples of implicit layers \cite{el_ghaoui_implicit_2021}. Implicit deep learning is a novel field in which layers are formulated as satisfying implicit conditions (e.g. $\text{NN}(x,y) = 0$), rather than as an explicit relationship between input and output (e.g. $y = \text{NN}(x)$). The key insight in this work is that these novel layers can be used to encode the steady-state and network boundary conditions. These layers resolve the initialization challenges of data-driven surrogate models allowing for seamless integration into a system model.

The complete End-to-End (E2E) integration of ML surrogates is a seldom-discussed step in other works and is one of the main focuses of this work. The contributions target the entire data-driven surrogate pipeline, including development, training, initialization, and integration into existing transient simulation workflows as shown in Fig. \ref{fig:ML_C}. 
Further, we aim to build tools that enable system operators to evaluate dynamic phenomena in grids with significant participation of IBRs that employ heterogeneous control techniques and structures.

\begin{figure}[t]
    \centering
    \includegraphics[width=1.0\columnwidth]{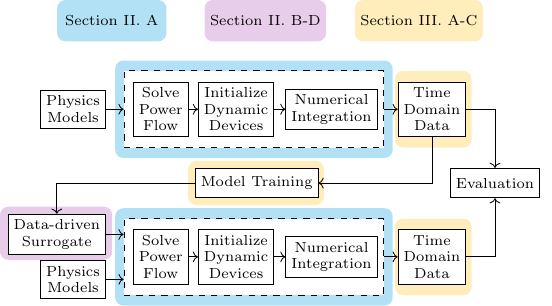}
    \caption{An overview of the proposed methodology. The different colors correspond to the focus of each subsection of the methodology.}
    \label{fig:methodology}
\end{figure}

The E2E methods developed in this manuscript are shown at a high level in Fig. \ref{fig:methodology} and consist of the following steps: (1)~A simulation is built and solved using physics-based models and numerical integration to generate time domain trajectories. (2)~The time domain simulation data is used as the ``ground truth'' data to train a data-driven surrogate model. (3)~The data-driven model is used along with physics-based models to build and solve a new simulation. (4)~The two datasets are compared on the basis of accuracy and simulation speed. 

The main contributions of this paper are: 
\begin{itemize}
    \item A data-driven surrogate structure, consisting of both DEQ and neural ODE layers (Section \ref{section:surrogate}), which can be integrated into time domain simulation procedures. 
    \item An E2E training, validation, and integration procedure for data-driven surrogate models in time domain simulations including EMT dynamics.
    \item Numerical results for a medium-size case study demonstrating the implementation of the proposed approach in an existing simulation platform (Section \ref{section:casestudy}).
\end{itemize}

\section{Surrogate Structure}
\label{section:surrogate}

\begin{figure*}[t]
    \centering
    \includegraphics[width=1.0\textwidth]{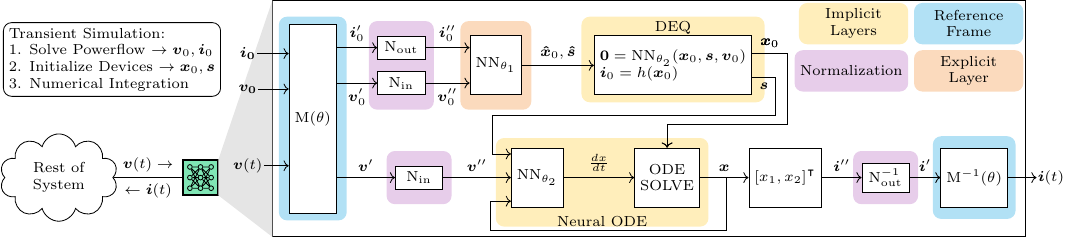}
    \caption{The proposed surrogate model. The blocks are color-coded according to their function---a key novelty lies in the combination of DEQ and NODE implicit layers with shared parameters.}
    \label{fig:surrogate}
\end{figure*}

The block diagram of the surrogate model is shown in Fig. \ref{fig:surrogate}. The proposed surrogate design is informed by the requirements of the procedure followed by standard time domain simulation tools described in Section \ref{subsection:simulation procedure}. 

\subsection{Power Systems Time Domain Simulation Procedure}
\label{subsection:simulation procedure}

A time domain simulation requires the specification of two components: (1)~a system model, including the differential algebraic equations that describe the system's physics and controls, and (2)~a time-stepping (or integration) algorithm \cite{lara_revisiting_2023}. In this work, we assume the system model consists of a set of time-invariant differential equations:
\begin{subequations}
\begin{align}
    \frac{d\boldsymbol{x}_{sys}}{dt} &= f_{sys}(\boldsymbol{x}_{sys}, \boldsymbol{y}_{sys}, \boldsymbol{s}_{sys}, \boldsymbol{\eta}_{sys}) \label{eq:sim1}\\
    \frac{d\boldsymbol{y}_{sys}}{dt} &= g_{sys}(\boldsymbol{x}_{sys}, \boldsymbol{y}_{sys}, \boldsymbol{\psi}_{sys}) \label{eq:sim2}
\end{align}
\end{subequations}
\noindent where ${\boldsymbol{x}_{sys}}$, $f_{sys}(\cdot)$, $\boldsymbol{s}_{sys}$, and $\boldsymbol{\eta}_{sys}$ represent, respectively, the states, equations, references, and parameters of the devices (e.g., inverters, machines, loads, surrogate models, etc.). The circuit dynamics of the network are represented as the sub-system $\boldsymbol{y}_{sys}$ and $g_{sys}(\cdot)$ with network parameters $\boldsymbol{\psi}_{sys}$.

Given the system model \eqref{eq:sim1}--\eqref{eq:sim2}, we define the procedure to execute a time domain simulation as the following three steps: 
\begin{enumerate}
    \item Solve the network model power flow $g_{sys}(\cdot)$, resulting in an initial condition for the network states $\boldsymbol{y}_{sys}^0$.
    \item Based on the power-flow solution ($\boldsymbol{y}_{sys}^0$), solve the steady-state equations for each device to determine $\boldsymbol{x}_{sys}^0$ and obtain the system-level steady-state.
    \item Apply a disturbance and numerically integrate, starting from the initial conditions and using a stepping algorithm. 
\end{enumerate}

The way in which each device interfaces with the network, and therefore the rest of the system, is also an important feature of the simulation definition. We assume a current balance framework wherein the input to each device is the voltage at the point of common coupling and the output is an injected current \cite{milano_power_2010}. With this formulation, a dynamic surrogate model can be written as: 
\begin{subequations}
\begin{alignat}{1}
    \dot{\boldsymbol{x}} &= f(\boldsymbol{x}, \boldsymbol{s}, \boldsymbol{v}, \boldsymbol{\eta}) \label{eq:device1}\\  
     \boldsymbol{i} &= h(\boldsymbol{x}, \boldsymbol{v}, \boldsymbol{\eta}) \label{eq:device2} 
\end{alignat}
\end{subequations}
\noindent where $\boldsymbol{x}$ are the dynamic states, $\boldsymbol{v}$ is the voltage input to the model, $\boldsymbol{i}$ is the model output current, $\boldsymbol{\eta}$ are parameters of the model, and $\boldsymbol{s}$ are references which are determined during initialization. Function $f$ represents the model's differential equations, and $h$ is the mapping from the dynamic states to the output current. 

\noindent \textbf{Definition} (Initial Network Interface Requirements). Given $\boldsymbol{v}_0, \boldsymbol{i}_0$ as the initial voltage and current from the power-flow solution, we define that a surrogate model \textbf{\textit{satisfies the network interface requirements}} if the following conditions are met:
\begin{subequations}
\begin{alignat}{1}
 &\boldsymbol{0} = f(\boldsymbol{x}_0, \boldsymbol{s}, \boldsymbol{v}_0, \boldsymbol{\eta})\label{eq:interface1} \\ 
 &\boldsymbol{i}_0 = h(\boldsymbol{x}_0, \boldsymbol{v}_0, \boldsymbol{\eta}) \label{eq:interface2}
\end{alignat}
\label{eq:interface}
\end{subequations} 
\noindent This definition implies that for a surrogate model to satisfy the initial network interface requirements, the differential terms must be initialized to a steady state while also matching the initial voltage and current values of the power flow solution.

This requirement is critical because if the surrogate is not in steady-state, it is not possible to find a full system condition that is in steady-state, which is typically the desired condition before a perturbation is applied. Furthermore, if the surrogate does not match the power flow quantities, the devices outside of the surrogate model have to compensate to achieve system balance leading to a different initial state than intended.

\subsection{Network Interface}
To reintegrate a surrogate into a simulation, the input and output network quantities ($\boldsymbol{v}$ and $\boldsymbol{i}$) need to be in the system reference frame. Internally, these quantities are first transformed to a local reference frame ($\boldsymbol{v}'$ and $\boldsymbol{i}'$) and then normalized ($\boldsymbol{v}''$ and $\boldsymbol{i}''$) in order to enhance the performance of the data-driven models. Without a local reference frame transformation, the solution of the surrogate model would change depending on the arbitrary definition of the system reference voltage. We define a local reference frame rotation to align one component of the input voltage to zero.
\begin{equation}
\label{eq:ref_surrogate}
\begin{aligned}
        \boldsymbol{v}'  = M(\theta)\boldsymbol{v}, \quad
        \boldsymbol{i}' = M(\theta) \boldsymbol{i}, \quad
      \theta = \arctan\left(\frac{v_{i}}{v_{r}}\right)
\end{aligned}
\end{equation}
where $M(\theta) =\begin{bmatrix} \text{sin}(\theta) & -\text{cos}(\theta) \\ \text{cos}(\theta) & \text{sin}(\theta)\\ \end{bmatrix}$, $\boldsymbol{v} = [v_r, v_i]^\intercal$, $\boldsymbol{i} = [i_r, i_i]^\intercal$, and $\theta$ is the time-varying angle of the input voltage in the system reference frame. This technique is analogous to the reference frame transformations which are common in many inverter and machine models \cite{park_two-reaction_1929, lara_revisiting_2023_tpwrs}.

ML workflows include some data preprocessing steps so that inputs to neural networks are normalized (e.g. bounded to $[-1, 1]$). Min-max normalization is included to normalize both the inputs and outputs based on the range of values seen in the training data:
\begin{equation}
\begin{aligned}
    \boldsymbol{v}'' = N_{\text{in}}(\boldsymbol{v}')= -1 + \frac{2(\boldsymbol{v}' - \boldsymbol{v}'_{min})}{\boldsymbol{v}'_{max} - \boldsymbol{v}'_{min}} \\
    \boldsymbol{i}'' = N_{\text{out}}(\boldsymbol{i}') = -1 + \frac{2(\boldsymbol{i}' - \boldsymbol{i}'_{min})}{\boldsymbol{i}'_{max} - \boldsymbol{i}'_{min}}
\end{aligned}
\end{equation}
\noindent where $\boldsymbol{v}'_{min}$ and $\boldsymbol{v}'_{max}$ are vectors of the minimum and maximum values respectively of each element of $\boldsymbol{v}'$ across the entire training dataset. The inverse of the normalization is included at the output of the model. Care should be taken to ensure the ranges of values in the validation and test datasets are similar to those expected to be found in both the training dataset and in practical usage. 

\subsection{Initialization}
The initialization portion of the proposed surrogate is a key differentiating feature with respect to prior work. As mentioned in Section \ref{subsection:simulation procedure}, the network interface requirements \eqref{eq:interface1}-\eqref{eq:interface2} are explicitly considered during initialization. In \cite{xiao_feasibility_2022}, the initial conditions for the neural ODE are calculated by passing the power flow quantities to an \textit{explicit} feed-forward neural network. However, there is no guarantee that the resulting initial conditions satisfy the network interface requirements. In contrast, the initialization of the proposed surrogate takes place through an \textit{implicit} layer---a DEQ layer to be precise---such that the forward pass of the model ensures the relevant conditions are met. This feature implies that even the \textit{untrained} surrogate matches the network interface requirements (see Fig. \ref{fig:expository}). Unless the initial conditions of a model can be calculated precisely in closed form (which is sometimes possible with physics-based models), the initialization process is fundamentally implicit. Therefore it is reasonable to model using an implicit ML layer. This approach embeds the network interface requirements into the structure of the surrogate, and satisfying the requirements is not reliant on the training process.  

The initialization portion of the surrogate is defined as: 
\begin{subequations}
\begin{alignat}{1}
&(\boldsymbol{i}_0,\boldsymbol{v}_0, \boldsymbol{p}) \rightarrow [\boldsymbol{x}_0^\intercal, \boldsymbol{s}^\intercal]^\intercal \\ 
&\boldsymbol{p} = [\boldsymbol{p}_1^\intercal, \boldsymbol{p}_2^\intercal]^\intercal \\
\text{s.t.} \nonumber \\
 &\boldsymbol{0} = \text{NN}_{\boldsymbol{p}_2}(\boldsymbol{x}_0, \boldsymbol{s}, \boldsymbol{v}_0'') \label{eq:init_surr_implicit_b} \\ 
 &\boldsymbol{i}_0'' = [x_{0_1}, \; x_{0_2}]^\intercal \label{eq:init_surr_implicit_c}  \
\end{alignat}
\label{eq:init_surr_implicit}
\end{subequations}
\noindent where the system of equations are solved numerically with the initial condition:
\begin{equation}
\label{eq:init_surr_explicit}
[\boldsymbol{\hat{x}}_0^\intercal, \;  \boldsymbol{\hat{s}}^\intercal]^\intercal = \text{NN}_{\boldsymbol{p}_1}(\boldsymbol{i}_0'', \boldsymbol{v}_0'')
\end{equation}

 The initialization consists of an explicit prediction from the first neural network \eqref{eq:init_surr_explicit} followed by a ``correction'' by the implicit layer \eqref{eq:init_surr_implicit_b}-\eqref{eq:init_surr_implicit_c}. The equations that are solved in the implicit layer guarantee that the surrogate satisfies the network interface requirements. The explicit prediction layer is given by $\text{NN}_{\boldsymbol{p_1}}$---a densely connected feed-forward neural network with parameters (weights and biases) $\boldsymbol{p_1}$ and the initial operating point as the input:
\begin{subequations}
 \begin{align}
          [\boldsymbol{i}_0''^\intercal, \boldsymbol{v}_0''^\intercal]^\intercal =& \mathbf{z^1_0} \\ 
          \mathbf{z^1_{k+1}} = \sigma(&\mathbf{W^1_{k+1}z^1_k} + \mathbf{b^1_{k+1}})\; \ \forall k = 0,  ... , K-1  \\
          [\boldsymbol{\hat{x}}_0, \boldsymbol{\hat{s}}] = \sigma(&\mathbf{W^1_{K+1}z^1_K} + \mathbf{b^1_{K+1}}) 
 \end{align}
 \end{subequations}
 The number of hidden layers $K$ and the width of the hidden layers $\text{dim}(\mathbf{b_k}) \ \forall k=1, ..., K-1$ are chosen as hyperparameters of the training problem. The output dimension $\text{dim}\left([\boldsymbol{\hat{x}}_0, \boldsymbol{\hat{s}}]\right)$ depends on the number of states, which is also a hyperparameter. In summary, the initialization mapping of the surrogate is given by an implicit-explicit layer similar to the layer proposed in \cite{pal_mixing_2022} as an isolated deep learning model for non-power systems applications.

The explicit layer is important in the overall structure of the surrogate as it allows the surrogate to generalize over different operating points (contingent on good training). The explicit layer is trained to give a good initial guess so that $[\boldsymbol{\hat{x}}_0, \boldsymbol{\hat{s}}] \approx [\boldsymbol{x}_0,\boldsymbol{s}]$. This implies that a different operating point $\boldsymbol{i}_0$, $\boldsymbol{v}_0$ will yield different initial conditions $\boldsymbol{x_0}$ for the neural ODE and therefore a different dynamic response. In comparison, prior works have dealt with multiple operating points by defining the data-driven states relative to the steady state \cite{vorwerk_using_2022}, implying that the dynamic response of the model will be the same regardless of the operating point. This assumption is not always true. For instance, an inverter operating without any headroom will behave much differently from an inverter with sufficient headroom when faced with an increase in load.

\subsection{Differential Equations}
The dynamic portion of the proposed surrogate ($f$) is given by a neural ODE \cite{chen_neural_nodate} wherein the derivatives of the hidden states are defined as the output of a neural network: 
\begin{equation}
\begin{aligned}
f(\boldsymbol{x}, \boldsymbol{s}, \boldsymbol{v}, \boldsymbol{p}) :&= \text{NN}_{\boldsymbol{p}_2}(\boldsymbol{x},\boldsymbol{s},\boldsymbol{v}) \\
\end{aligned}
\end{equation}
The inputs to the neural ODE include the dynamic states ($\boldsymbol{x}$), the references which are determined during initialization ($\boldsymbol{s}$), and the time-varying exogenous input ($\boldsymbol{v}$) which comes from the interface with the rest of the system. The neural ODE neural network ($\text{NN}_{\boldsymbol{p}_2}$) is also a full-connected architecture with parameters $\boldsymbol{p}_2$: 
\begin{subequations}
\label{eq:nn2}
 \begin{align}
          [\boldsymbol{x}^\intercal, \boldsymbol{s}^\intercal, &\boldsymbol{v}^\intercal]^\intercal = \mathbf{z^2_0} \\ 
          \mathbf{z^2_{k+1}} &= \sigma(\mathbf{W^2_{k+1}z^2_k} + \mathbf{b^2_{k+1}})\; \forall k = 0,  ... , K-1  \\
          \dot{\boldsymbol{x}} &= \mathbf{W^2_{K+1}z^2_K}  \label{eq:unconstrained}
 \end{align}
 \end{subequations}
The key difference is that the output layer is not bounded by an activation function and does not include a bias. From a numerical perspective, the neural ODE describes a non-linear relationship which defines the derivatives of the states and therefore the neural ODE architecture is compatible with any numerical integration algorithm. 

The final portion of the model is the mapping from the dynamic states ($\boldsymbol{x}$) to the outputs which interface with the rest of the system. In this case, the simplest possible option is chosen---the first two dynamic states are selected to represent the output and the inverse of the normalization and reference frame transformations are applied to obtain the output current in the system reference frame and units:
\begin{equation}
\label{eq:h_surrogate}
h(\boldsymbol{x},\boldsymbol{p}) := M^{-1}(\theta)N_{\text{out}}^{-1}([x_1, \; x_2]^\intercal) = \boldsymbol{i}
\end{equation}
\noindent The complete surrogate model (Fig. \ref{fig:surrogate}) is given by \eqref{eq:ref_surrogate} -- \eqref{eq:h_surrogate}. 

\begin{figure}
\centering
\subfigure[Surrogate before training.]
{
\includegraphics[width=0.95\linewidth]{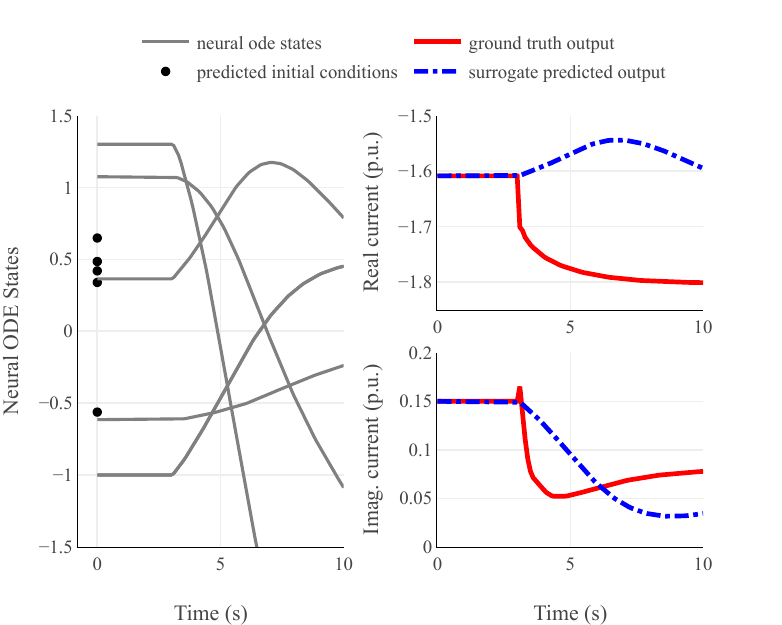}
\label{fig:expository_before}
}
\subfigure[Surrogate after training.]
{
\includegraphics[width=0.95\linewidth]{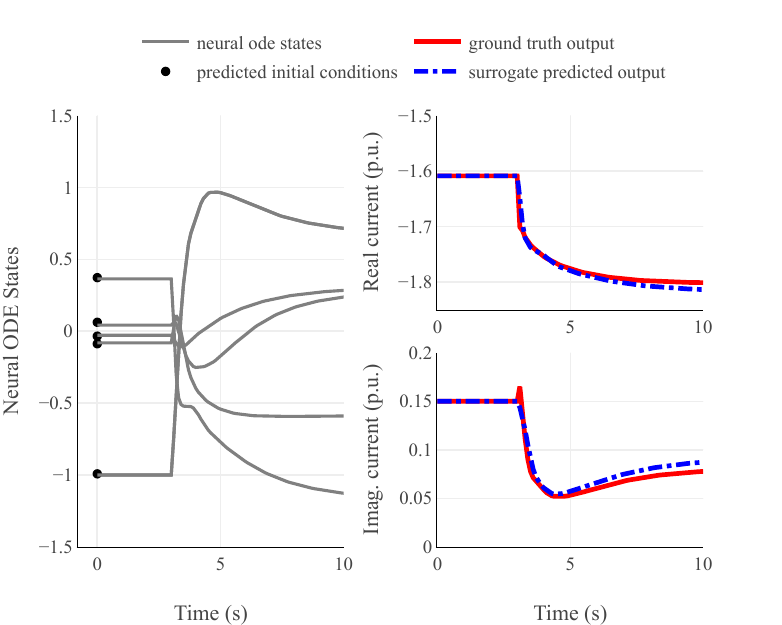} 
\label{fig:expository_after}
}
\caption{An expository result---(a) and (b) show the same quantities for the surrogate before and after training respectively. The left side shows the trajectories for the hidden states ($\boldsymbol{x}$) along with the predicted initial conditions ($\boldsymbol{\hat{x}}_0$) shown as black circles. The right side shows the model output (real and imaginary current).}
\label{fig:expository}
\end{figure}

One way to visualize the novel properties of the proposed surrogate is to consider the behavior of an \textit{untrained} surrogate model. While this is not common practice in a traditional ML model, in this case the untrained behavior helps demonstrate the parts of the model performance which rely on training and the parts which are inherent due to the structure. A sample result is shown in Fig. \ref{fig:expository} for the proposed surrogate, both before (Fig. \ref{fig:expository_before}) and after (Fig. \ref{fig:expository_after}) training. On the left side are the dynamic states of the neural ODE and on the right side are the outputs of interest (real and imaginary current). The predicted initial states ($\boldsymbol{\hat{x}}_0$) are plotted as circles at time zero. As expected, before any training, there is a significant difference between the predicted and actual initial states $\boldsymbol{\hat{x}}_0$ and $\boldsymbol{x}_0$, and the dynamic response does not match the ground truth. However, the hidden states are in steady state (flat line before the perturbation is applied) and the initial outputs match exactly the ground truth data, precisely because both these constraints are encoded into the DEQ layer. After the surrogate is trained (Fig. \ref{fig:expository_after}), the prediction of the initial conditions is much closer, as is the dynamic response.

\section{Training Methodology}
\label{subsection:training}
While the parameters of a physics-based model can often be derived from the physical system itself, the parameters of the proposed data-driven models must be learned from data during the training process. In practice, this means that the surrogate structure described must be fully differentiable so that the gradient of each parameter with respect to the loss can be calculated during training. The details of the training process are discussed in the following section.  

\subsection{Data Generation} The dataset used for training and evaluation of the model, $\mathcal{D}$, consists of $n$ pairs of input and output time-series trajectories where each trajectory contains $m$ time points:   
\begin{equation}
\mathcal{D} = \{\{(\boldsymbol{v}_{(i,j)},\boldsymbol{i}_{(i,j)})\}_{i=1}^{m}\}_{j=1}^n 
\end{equation}
For the purpose of this paper, the input is the bus voltage at the point of common coupling ($\boldsymbol{v} = [v_r, v_i]^\intercal \in \mathbb{R}^2$) and the output is the injected current from the surrogate ($\boldsymbol{i} = [i_r, i_i]^\intercal \in \mathbb{R}^2$). The full dataset of $n$ trajectories is randomly split into train, validation, and test datasets.

 In this work all training data is generated from numerical simulations of the full system model including EMT dynamics as employed in \cite{henriquez2020grid} (see Section \ref{subsection:computational} for implementation details). The dataset consists of the network quantities (voltages, currents) at the point of common coupling between the surrogate portion and the remainder of the system. Generating data by simulating perturbations of interest in the full system is an easy way to ensure that the relevant dynamics that exist in the test dataset are captured during training, even though the train and test datasets consist of distinct trajectories that are generated independently. While the black-box structure of the surrogate lends itself to applications in which an accurate physical model is not known and only data is available (e.g. a real distribution feeder with measurement data at the sub-station transformer), the extension to measurement data is beyond the scope of this work.

\subsection{Loss Function}
The purpose of the loss function is to quantify the performance of the surrogate model during training so that the parameters can be iteratively updated (trained), thereby improving the performance. The loss function is given by:
\begin{equation}
\begin{aligned}
\mathcal{L}(\boldsymbol{p}_1, \boldsymbol{p}_2) =& \;(1-\alpha)\mathcal{L}_\text{init.} + \alpha \mathcal{L}_\text{dyn.} \\ 
\mathcal{L}_\text{init.} =& \; \text{RMSE}\left([\boldsymbol{x}_0^\intercal, \boldsymbol{s}^\intercal], [\boldsymbol{\hat{x}_0}^\intercal, \boldsymbol{\hat{s}}^\intercal]\right)  \\   
\mathcal{L}_\text{dyn.} =& \; \text{RMSE}\left([\boldsymbol{i}_{(1)}^\intercal \; ... \; \boldsymbol{i}_{(m)}^\intercal], [\boldsymbol{\hat{i}}_{(1)}^\intercal \; ... \; \boldsymbol{\hat{i}}_{(m)}^\intercal]   \right)
\end{aligned}
\label{eq:loss}
\end{equation}
where RMSE is the well known root mean square error:
\begin{equation*}
    \text{RMSE}(\boldsymbol{x},\boldsymbol{\hat{x}}) =\sqrt{\frac{\sum_{i=1}^{n}\left(x_i - \hat{x}_i\right)^2}{n}}
\end{equation*}
$\boldsymbol{p}_1$ and $\boldsymbol{p}_2$ are the parameters of the explicit initialization prediction layer and neural ODE, respectively. The initialization loss, $\mathcal{L}_\text{init.}$, quantifies how well the explicit 
initial condition prediction layer matches the actual initial conditions. The dynamic loss, $\mathcal{L}_\text{dyn.}$, quantifies how well the predicted output trajectories match the ground truth data. Both $\mathcal{L}_\text{init.}$ and $\mathcal{L}_\text{dyn.}$ depend on both sets of parameters, therefore all parameters are trained simultaneously according to the total loss. $\alpha$ is a tradeoff factor between the initialization and dynamic loss and is tuned as a hyperparameter of the training.  

\subsection{Training Algorithm}

The key challenge in training the proposed surrogate is to maintain the convergence of both the DEQ and neural ODE layers throughout training. To the best of the authors' knowledge, this work is unique in combining a DEQ layer and neural ODE with shared parameters, hence there are unique training challenges. In \cite{kim_stiff_2021}, the authors note the importance of equation scaling to promote stability while training neural ODEs. For DEQs, it is desirable for the neural network to be constrained in order to promote stability \cite{bai_deep_nodate}. In this case, however, the output layer of the DEQ neural network must be unconstrained because it represents the derivative of the model states which should be allowed to take on all real values (\ref{eq:unconstrained}). 

In order to address the convergence of the DEQ layer, the training algorithm (Algorithm 1) includes a pre-processing step to ensure that the starting random initialization of the neural networks does not lead to divergent behavior when the training begins. To achieve this, the neural networks are randomly initialized and the convergence of the DEQ layer is checked for each entry in the training set. This ensures that for the first training iteration the surrogate will converge and we can calculate a meaningful value for both $\mathcal{L}_\text{init.}$ and $\mathcal{L}_\text{dyn.}$. For the actual training portion, the loss is calculated stochastically for a single trajectory for each training iteration and the parameters are updated according to the first-order Adam algorithm \cite{kingma_adam_2017}. In order to prevent over fitting, the loss on a validation set is checked at regular intervals during training. The training ends when a maximum number of iterations is met. The hyper-parameters of the training process (Table \ref{tab:hyperparameters}) are tuned and selected based on the best performance on the validation dataset, and the final performance results are evaluated on an unseen test dataset.

\begin{algorithm}
\caption{Surrogate Training Procedure}\label{alg:cap}
\begin{algorithmic}
\Require $\mathcal{D}_{\text{train}} = (u_{(i,j)},y_{(i,j)})$, $k_{\text{max}}$ 
\While{DEQ does not converge for all of $\mathcal{D}_{\text{train}}$ }
    \State Randomly initialize $p_1$, $p_2$
\EndWhile 
\While{$k<k_{\text{max}}$}
\State Choose a perturbation at random to calculate loss.
\State update $p_1$, $p_2$ according to Adam.
\EndWhile
\end{algorithmic}
\end{algorithm}

\section{Results}
\label{section:casestudy}
\begin{figure}[t]
    \centering
    \includegraphics[width=0.485\textwidth]{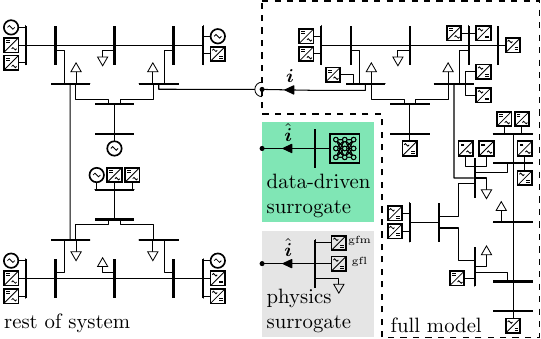}
    \caption{The case study system. The right side shows the full-order model (unshaded) and the two surrogate models (shaded) which can each be connected to the remainder of the system. The performance criteria is based on how well the predicted current ($\hat{\boldsymbol{i}} = [\hat{i}_r, \hat{i}_i]^\intercal$) matches the ground-truth result ($\boldsymbol{i} = [i_r, i_i]^\intercal$).}
    \label{fig:36bus}
\end{figure}
\subsection{Computational Set Up}
\label{subsection:computational}

All of the code for generating data, training the surrogates, and evaluating the performance is open source and written in the Julia programming language. Julia is chosen for this work for its unique blend of high performance and ease of use. \texttt{PowerSimulationsDynamics.jl}---a package for simulating the transient response of power systems with large penetrations of IBRs---is used to generate the train, validation, and test datasets \cite{lara2023powersimulationsdynamics}. The surrogate model shown in Fig. \ref{fig:surrogate} is built using the packages of the Julia Scientific Machine Learning ecosystem \cite{noauthor_sciml_nodate} so that the entire process of initializing and numerically integrating to solve for time domain trajectories is fully differentiable. After training, the composable nature of Julia along with the design of the surrogate make it possible to re-integrate trained data-driven models into \texttt{PowerSimulationsDynamics.jl} by defining a new device model. For the full details of the packages used and the experiments presented in this section, the reader is directed to the paper repository\footnote{https://github.com/m-bossart/PowerSystemNODEs}. 
 
\subsection{Case Study}
\label{subsection:casestudy}

\begin{table}[t]
\centering
\caption{Summary of dynamic models.}
\begin{tabular}{l | l | c}
Component & Model  & Dynamic States \\
\hline
\hline
Transmission lines & $\pi$-model & 6 \\
Machines  & GENROE & 4 \\
Excitation System & Type II AVR & 4 \\
Turbine Governor & Type II & 1 \\
Inverter A & Grid-following \cite{kenyon_open-source_2021} & 16 \\ 
Inverter B & Grid-forming - Droop \cite{darco_virtual_2015} & 15 \\ 
\end{tabular}
\label{tab:dynamic_models}
\end{table}

In this section, the training and evaluation of the proposed surrogate are demonstrated on the synthetic 36-bus power system shown in Fig. \ref{fig:36bus}. The system is modeled in detail such that the inverter inner control loops and line dynamics are not neglected. The dynamic models are summarized in Table \ref{tab:dynamic_models} and the full detailed descriptions can be found in the \texttt{PowerSimulationsDynamics.jl} 
 documentation \cite{lara2023powersimulationsdynamics}. The 36-bus system as constructed contains 719 dynamic states in total, making it a medium-sized system with non-trivial computational burden to solve.

\begin{figure}[t]
    \centering
    \includegraphics[width=0.9\columnwidth]{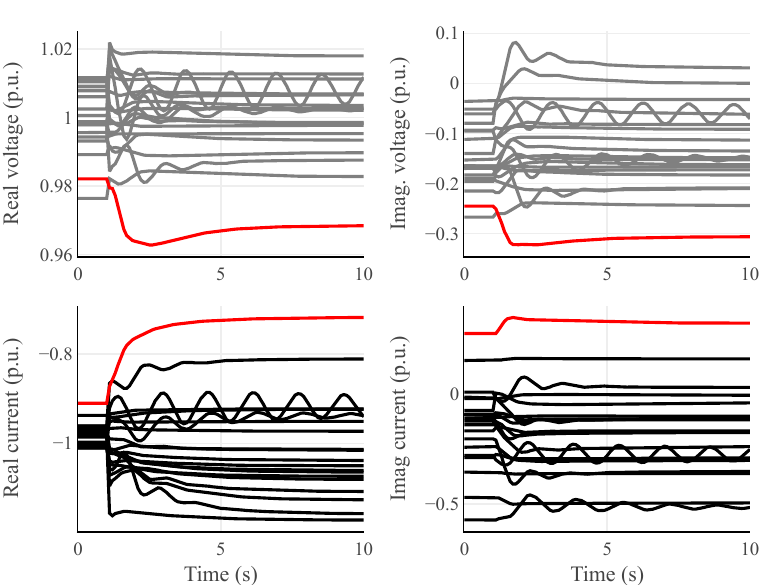}
    \caption{The trajectories of the train dataset. The top row shows the model inputs (voltage) in gray, and the bottom row shows the model outputs (current) in black. The red traces correspond to the largest single load step (85\% step) in the train dataset.}
    \label{fig:datasets}
\end{figure}

The system is simulated across a range of operating conditions and perturbations to generate the ground truth trajectories that make up the train, validation, and test datasets. For each trajectory, the operating point of the base system is changed by randomly scaling the set points between a fixed range for each generator and load that is part of the external system (i.e. the left side of Fig. \ref{fig:36bus}). The generator voltage set points are randomly sampled from a uniform distribution $[0.96, 1.04]$, the generator active power references are randomly sampled from $[0.0, 1.0]$, and the load active and reactive powers are scaled by a factor randomly sampled from $[0.5, 1.5]$ as in 
\cite{xiao_feasibility_2022}. The perturbations for the base study consist of step load changes wherein the active and reactive power of a randomly selected load are both scaled by a factor sampled from $[0.0, 2.0]$. Each simulation is run for a total of 10 seconds with the perturbation applied at 1 second. The recorded states are the real and imaginary voltage and real and imaginary current at the point of common coupling between the portion of the system which will be modeled with the surrogate and the remainder of the system. For each trajectory, 100 data points are saved at a linear spacing of 0.1 seconds ($m=100$). The trajectories of the train dataset are shown in Fig. \ref{fig:datasets}.

The performance of the proposed surrogate is compared against a benchmark physics-based surrogate model. The physics-based surrogate is based on the common practice of aggregating multiple similar devices by representing the collective behavior with a single model. The portion of the system to be replaced with a surrogate contains two types of inverters and one type of load, therefore the physics-based surrogate is comprised of two aggregate inverter models and one aggregate load model located at a single bus (the dynamic equations of the aggregate models and the individual device models are the same). The initial parameters of the physics-based surrogate model are determined by a base-power weighted average of the same parameter for the individual devices. The base-power of the aggregate devices is the sum of the individual base powers. While this set of parameters is a reasonable starting point, there is no guarantee that this particular parameterization best captures the behavior of the true model across the test dataset. Therefore, in order to make a more fair comparison, the physics-based surrogate model is also trained using the Adam algorithm on the same training dataset. If all of the physics-based parameters are included in the training, the underlying model is easily perturbed to an unstable parameterization, therefore only a key subset of the parameters are learned. These include the proportion of load modeled as constant impedance/constant power/constant current, the proportional and integral gains for active and reactive power control in the grid-following inverter, and the active and reactive power droop coefficients of the grid-forming inverter. Both the ``untrained'' and ``trained'' physics-based surrogates are included as points of comparison in the following results.

Hyperparameter tuning is conducted using a basic random search algorithm in order to explore a range of options as summarized in Table \ref{tab:hyperparameters}. Most of the relevant hyperparameters are related to the size of the neural networks within the data-driven surrogate. The structure of the physics-based surrogate does not have any hyperparameters, however the initial step norm for the Adam algorithm is also included as a tunable hyperparameter. For the data-driven surrogate a total of 200 random hyperparameter combinations are trained and the top performing set of hyperparameters on the validation set are selected. Given that the physics-based model has only one hyperparameter, only 33 options are evaluated. 

\begin{table}[t]
\centering
    \setlength\tabcolsep{2pt} 
\caption{Summary of tunable hyperparameters considered for data-driven and physics-based surrogates.}
\begin{tabular}{c |c| l  |c| c | c}
& \multirow{2}{1.1cm}{Parameter} & \multirow{2}{1.2cm}{Description} & Lower  & Upper  & Chosen  \\
&  &  &   limit &  limit &  Value \\
\hline
\noalign{\vskip 2mm}  
\hline

\multirow{8}{0.8cm}{data-driven param.}
& $n_{init}$ & number of hidden layers in $\text{NN}_1$ & 1 & 3 & 2\\
& $w_{init}$ & neurons in hidden layers of $\text{NN}_1$ &  3 & 23 & 17\\
& $n_{dyn}$ &  number of hidden layers in $\text{NN}_2$ & 1 & 3 & 1\\
& $w_{dyn}$ & neurons in hidden layers of $\text{NN}_2$ &  6 & 39 & 16\\
& $x_{dim}$ & hidden states &2 & 15 & 14\\
& $\log_{10}(\eta)$ & log of ADAM learning rate &-3.0 & -1.0 & -1.52\\
& $\alpha$  & scaling of init. and dyn. loss & 0.1 & 0.9 & 0.76\\
\hline
\noalign{\vskip 2mm}  
\hline
\multirow{2}{0.8cm}{physics param.}& \multirow{2}{*}{$\log_{10}(\eta)$} &  \multirow{2}{*}{log of ADAM learning rate} & \multirow{2}{*}{-7.0} & \multirow{2}{*}{-1.0} & \multirow{2}{*}{-7.0}\\
&  &  &  & & \\
\end{tabular}
\label{tab:hyperparameters}
\end{table}

\begin{figure}[t]
    \centering
    \includegraphics[width=0.48\textwidth]{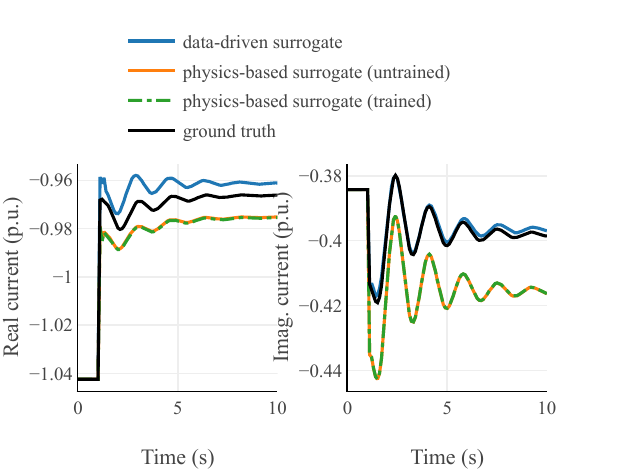}
    \caption{A single sample trajectory from the test dataset. The MAEs corresponding to this trajectory are highlighted in black in Fig. \ref{fig:accuracy}.}
    \label{fig:sample_trajectory}
\end{figure}

\begin{figure}[t]
    \centering
    \includegraphics[width=0.48\textwidth]{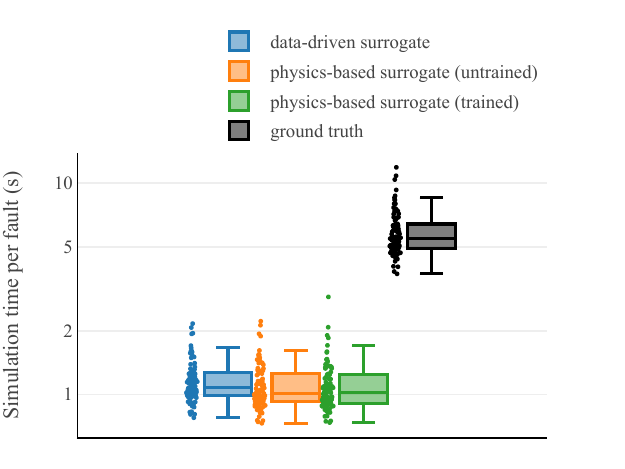}
    \caption{Simulation time comparison for the system of Fig. \ref{fig:36bus} with the full order model and each of the surrogate models. The times are plotted on a log scale, and the systems with surrogate models are approximately four times faster on average.}
    \label{fig:timing}
\end{figure}
\begin{figure}[t]
    \centering
    \includegraphics[width=0.48\textwidth]{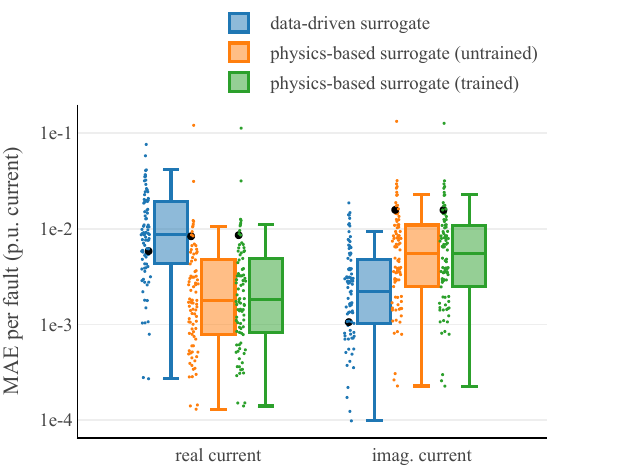}
    \caption{MAE for the data-driven and physics-based surrogates compared to the full order model. The error is plotted on a log scale, and each point represents the mean error for a single perturbation trajectory (90 data points total). The black dots correspond to the trajectory shown in Fig. \ref{fig:sample_trajectory}.}
    \label{fig:accuracy}
\end{figure}

After the hyperparameters are chosen, the surrogates are finally evaluated on the test dataset, which consists of 90 trajectories generated by the same process used for the train and validation datasets. All of the subsequent results are contingent on the encoding of the initial network interface requirements in the surrogate design, which allows for running system level simulations with the data-driven surrogate included. Figure \ref{fig:sample_trajectory} compares a single sample trajectory from the test dataset. For this trajectory, the oscillations of the ground truth data are captured with both types of surrogates, however, the imaginary current error is significantly lower for the data-driven surrogate. The training process results in negligible improvements for the physics-based surrogate, indicating that the starting parameters capture the behavior well and that the structure of the model limits the ability to improve the performance across the entire dataset through training. This single trajectory is not representative of the performance overall; for more meaningful results, the surrogates are compared on the basis of simulation speed (Fig. \ref{fig:timing}) and error relative to the full order detailed model (Fig. \ref{fig:accuracy}) across the entire test dataset. The level of accuracy or error that is ``acceptable'' is specific to the downstream analysis enabled by the simulation. In this work, we demonstrate the accuracy of the surrogate compared to the physics-based alternative without considering a specific workflow; expanding the metrics and quantifying the quality of the model based on various use cases is left to future work. 

Fig. \ref{fig:timing} shows the distribution of simulation times for the test dataset. The results are straightforward and expected---the simulation time is dramatically reduced for each of the surrogate models due to the significant reduction in total number of states. The exact speedup is dependent on the properties of the system model, including, importantly, the stiffness and the type of solver used. In this case study, the speed up is approximately a factor of four for the data-driven and physics-based surrogates.
One of the benefits of the proposed approach, to learn only part of a larger system, is the scalability.  For example, learning and integrating many relatively small sub-systems (e.g. individual distribution systems) into a single larger simulation offers the potential to scale to very large systems in a modular manner.  

Fig. \ref{fig:accuracy} shows the mean absolute error (MAE) of real and imaginary current for each trajectory in the test dataset. The overall accuracy of the data-driven and physics-based surrogate is similar; the data-driven surrogate has slightly lower average error for real current and slightly higher average error for imaginary current. In contrast to the physics-based surrogate, which starts with almost optimal parameterization, the data-driven surrogate requires training to give a reasonable output, but is able to learn the behavior without imposing a model structure \textit{a priori}.

Table \ref{tab:timing} compares the full order model with the two surrogates on the basis of number of parameters, number of states, and simulation time for the test dataset. While the physics-based surrogate dramatically reduces the number of states and parameters, the data-driven surrogate has \textit{more parameters} than the original model. The expressiveness of ML models in general is related to the large number of trainable parameters \cite{shen_differentiable_2023}; in fact, neural networks of sufficient size are a class of universal function approximators \cite{hornik_approximation_1991}. In the context of modeling dynamic systems, the flexibility and expressiveness of the data-driven model comes largely from the parameters and the representation of non-linear relationships between states rather than a large number of dynamic states. This distinction is key for reducing simulation time because the number of states largely determines the simulation time, whereas the number of parameters has a minimal impact. 

\begin{table}[t]
\centering
\caption{Comparison of model complexity in terms of number of states and number of parameters.}
\begin{tabular}{l c c c}
Model & States & Parameters & *Simulation time (s) \\

\noalign{\vskip 1mm}  
\hline
\noalign{\vskip 2mm}  
Full order model & 432 & \textasciitilde 600 &  5.86\\
Physics-based surrogate & 31 & 54 & 1.28  \\
Data-driven surrogate & 14 & 1322 & 1.23\\
\noalign{\vskip 2mm}  
\multicolumn{4}{l}{*Average simulation time across the test dataset (90 trajectories).} \\
\end{tabular}
\label{tab:timing}
\end{table}

\begin{figure}[t]
    \centering
    \includegraphics[width=0.48\textwidth]{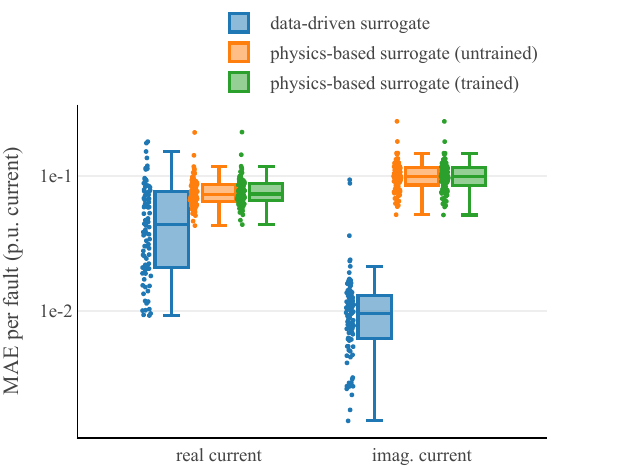}
    \caption{MAE for the out-of-sample test dataset. The perturbation for this dataset is tripping of half of the line connecting the surrogate with the remainder of the system.}
    \label{fig:accuracy_linetrip}
\end{figure}

In the results presented thus far, the train, validation, and test datasets are derived from the same processes for randomly changing the system operating point and applying a random load step. In order to test the generalization of the surrogate models, the models are also evaluated on a second out-of-sample test dataset. In this dataset, the operating point selection remains unchanged and the system is perturbed by tripping one half of the connecting line between the surrogate and the rest of the system. The error results are shown in Fig. \ref{fig:accuracy_linetrip}. In this case, the data-driven surrogate outperforms the physics-based surrogate in terms of imaginary current error. This result suggests that data-driven surrogate models with a large number of parameters can potentially generalize across a larger number of use cases compared to physics-based surrogates, which often require frequent re-parameterization \cite{noauthor_model_2018, noauthor_value_2015}. More work is needed to fully evaluate how the surrogate generalizes to unseen perturbations. 

It is important to note that the performance of physics-based and data-driven surrogates will be highly dependent on the underlying model, therefore one method will not always outperform the other. For example, for modeling many devices that are electrically close and very similar (e.g. individual wind turbines of a wind plant), physics-based aggregation is likely to provide sufficiently accurate results. For systems with a diversity of devices with varying control structures and parameters, such as are expected in the transition towards a carbon-free power system, the flexibility of data-driven approaches can be beneficial. 

\section{Conclusion}
\label{section:conclusion}

In this paper we define a set of initial network interface requirements and propose a data-driven surrogate structure which satisfies the requirements by design---ensuring that the trained models can be re-integrated into system-level simulations. The surrogate methodology is evaluated on a medium-sized power system with 719 states, and achieves similar accuracy (mean average current error of $8.3\times10^{-3}$) and speedup (4 times) as a physics-based surrogate. However, unlike the physics-based alternative, the proposed surrogate does not rely on detailed knowledge of the underlying device models, which is unlikely to be available for some applications. One avenue of future work is to apply the proposed methodology to distribution system modeling with high levels of DERs where incomplete network knowledge is particularly relevant. 

One of the downsides of data-driven methods as compared to physics-based approaches is the need for training, which can be a computationally expensive process. In this work, ground truth training data is generated through simulations of power system perturbations. An alternative approach is to define an exogenous perturbation and simulate the response of only the surrogate portion of the system. Such an approach can potentially improve the training efficiency; however, the exact form of the perturbing signal, such that the surrogate is accurate when exposed to real power system perturbations, is an open research question. 

Another key area of future work is elucidating the exact requirements for how general a data-driven surrogate must be in order to replace physics-based models in certain applications. For example, in this work the data-driven surrogate is shown to perform well across different operating conditions and types of perturbations. However, the surrogate cannot capture changes due to internal changes to the surrogate structure or the parameters of the ground truth model. Establishing benchmark problems will be key to determining the aspects of realistic performance that are needed to make data-driven approaches a practical alternative in the future. 

Finally, while this paper compares a purely data-driven and purely physics-based surrogate, future work should also consider hybrid models which can take advantage of the benefits of each method. The potential benefits of hybrid models include reduced data requirements for training, improved generalization, and improved interpretability \cite{shen_differentiable_2023}.

\ifCLASSOPTIONcaptionsoff
  \newpage
\fi



%
\bibliographystyle{IEEEtran}
\bibliography{NODE}
\end{document}